\documentclass{kluwer}    
\usepackage{graphics}

\newcommand{\degr}[1]{\ensuremath{#1^{\circ}}}         
\newcommand{\hi}{{\sc{Hi}}}                          

\newcommand{\Msun}{\mbox{\,M\ensuremath{_\odot}}}      
\newcommand{\kms}     {km~s$^{-1}$}                          
\newcommand{\Mhi}{\ensuremath{\mathrm{M_{\mbox{\scriptsize{\hi}}}}}} 

\newdisplay{guess}{Conjecture}

\begin{document}                                                                                   
\begin{article}
\begin{opening}         
\title{Neutral Hydrogen in Galaxy Groups}
\author{N. P. F. \surname{M$\mathrm{^c}$Kay}
}
\author{C. G. \surname{Mundell}}
\author{S. \surname{Brough}
}
\institute{Astrophysics Research Institute, Liverpool John Moores University}
\author{D. A. \surname{Forbes}}
\institute{Swinburne University of Technology}
\author{D. G. \surname{Barnes}}
\institute{The University of Melbourne}
  
\runningauthor{M$\mathrm{^c}$Kay et al.}
\runningtitle{The Role of \hi{} in the Evolution of Galaxy Groups}

\begin{abstract}
We present preliminary results from a study of the neutral hydrogen
(\hi{}) properties of an X-ray selected sample of nearby loose galaxy
groups.  This forms part of a multi-wavelength investigation (X-ray,
optical and radio) of the formation and evolution of galaxies within a
group environment.  Some initial findings of an ATNF Parkes Multibeam
wide-area neutral hydrogen (\hi{}) imaging survey of 17 nearby galaxy
groups include two new, potentially isolated clouds of \hi{} in the
NGC~1052 and NGC~5044 groups and significant amounts of \hi{} within
the group virial radii of groups NGC~3557 and IC~1459 - two groups
with complex X-ray structures that suggest they may still be in the
act of virialisation.  Here we present ATCA high-resolution
synthesis-imaging follow-up observations of the distribution and
kinematics of \hi{} in these four groups.
\end{abstract}
\end{opening}

\section{Introduction}  
Neutral hydrogen is a valuable tracer of galactic structure and
dynamics. It is the most spatially extended component of a galaxy's
disk, and is particularly sensitive to tidal interactions and
mergers between galaxies, retaining the imprint of an encounter out to
large galactic radii \cite{mpa+95}. It is therefore an ideal tool for probing
the morphology and dynamics of galaxies in groups, with the
aim of unravelling the processes driving galaxy evolution in a
group environment.

This study is part of the GEMS project - a multi-wavelength (X-ray,
optical and radio) investigation of the formation and evolution of
galaxies within groups We have conducted a wide-area ($\degr{5.5}
\times \degr{5.5}$ per target) neutral hydrogen~(\hi{}) imaging survey
of 17 nearby galaxy groups (distance $< 40$~Mpc) using the ATNF Parkes
Radiotelescope Multibeam System, at an angular resolution of
17~arcminutes.

Initial results of this survey include the detection of \hi{} within
the X-ray virial radius in the IC 1459 and NGC~3557 groups.
Additionally, new group members previously uncatalogued at optical
wavelengths were detected at the edges of the groups NGC~1052 and
NGC~5044.

High angular resolution ($\sim 35$~arcsec) synthesis followup
observations of a targeted 33~arcmin region in each of
these four groups were subsequently obtained, using
the Australia Telescope Compact Array, with the goal of determining
the precise location, distribution and kinematics of the \hi{} in
these regions.

\section{NGC~1052 Group}
\begin{figure}
  \begin{minipage}[h]{0.36\linewidth}
    \begin{center}
      \resizebox{\textwidth}{!}{
        \includegraphics{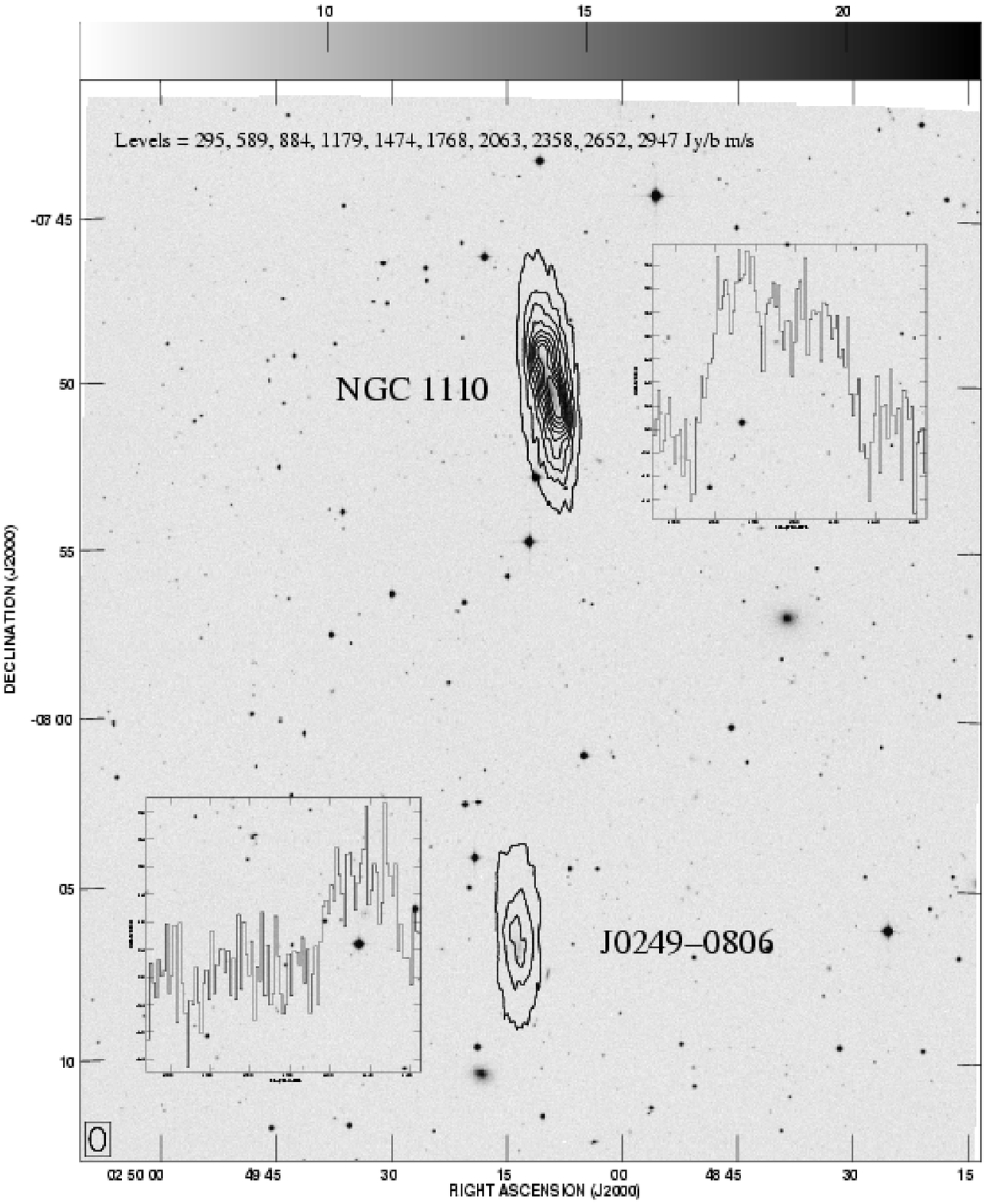}
      }
    \end{center}
  \end{minipage} \hfill
  \begin{minipage}[h]{0.36\linewidth}
    \begin{center}
      \resizebox{\textwidth}{!}{
        \includegraphics{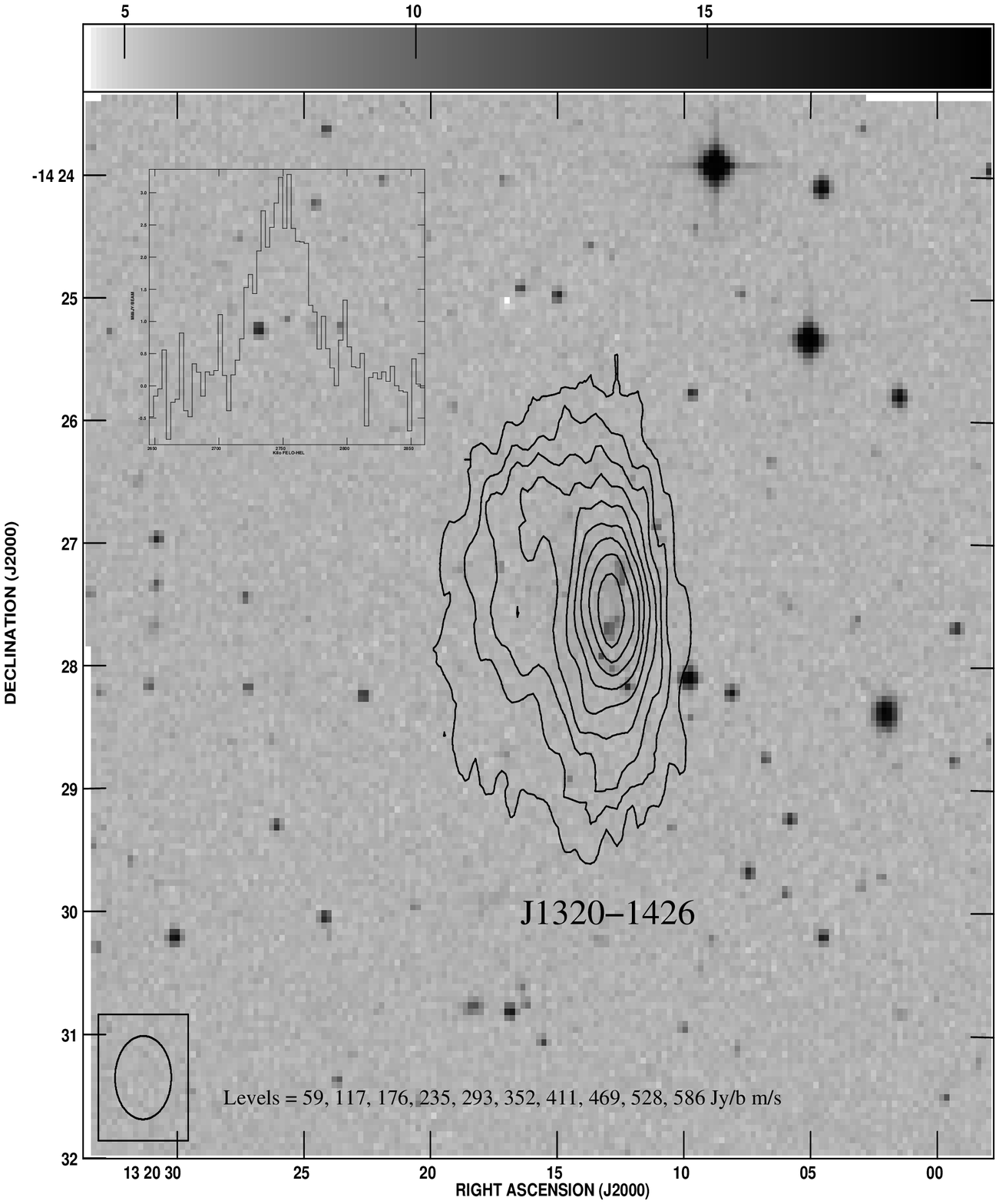}
      }
    \end{center}
  \end{minipage}

  \caption{{\em (Left)} NGC~1052 Group: \hi{} total intensity contours
  superimposed onto the DSS2-red image of NGC~1110 (top) and
  J0249-0806 (bottom), with ATCA spectra inset.
  {\em (Right)} NGC~5044 Group: \hi{} total intensity contours
  superimposed onto the DSS1 image of J1320-1426, with ATCA \hi{}
  spectrum inset.  } \label{fig:n1052_n5044}
\end{figure}

The loose group NGC~1052~\cite{gar93} has 14 catalogued member
galaxies and is located at a distance of 19.3~Mpc, with a systemic
velocity of 1438~\kms{}.  In addition to \hi{} detection of group
members, the Parkes wide-field observations of this group reveal \hi{}
emission from a source located $\sim\degr{2}$ (690~kpc) from the
centre of the group, and 14~arcmin (100~kpc) south of the edge-on
galaxy NGC~1110.  The ATCA followup observation, with spatial
resolution of $42''\times30''$ and spectral resolution of 3~\kms{},
confirms the \hi-emitting region to be $\sim 10 \times 50$~kpc in
extent, approximately 90~kpc south of, and spatially distinct from,
NGC~1110 (see Figure~\ref{fig:n1052_n5044}, left).  This object, to
which we refer as J0249-0806, has a mean velocity of 1450~\kms{}, a
derived \hi{} mass of $\Mhi~=~1.2~\times~10^9$~\Msun{} and a mean
$\sigma_{\mathrm{HI}} = 3$~\Msun~pc$^{-2}$
(cf. $\sigma_{\mathrm{HI}} \mathrm{\scriptstyle (LSB gal)} \sim
2$~\Msun~pc$^{-2}$ \cite{dmv}).
The DSS2-red optical image at the position of J0249-0806 shows
uncatalogued faint emission spatially coincident with the \hi{} gas
(see Figure~\ref{fig:n1052_n5044}, left), and optical data now in hand
will allow us to better determine the nature of his object.

\section{NGC~5044 Group}
The NGC~5044 group has 9 catalogued member galaxies~\cite{gar93}, a
systemic velocity of 2379~\kms{} and a distance of 19.3~Mpc.  A
previously uncatalogued source of \hi{} emission was discovered by the
Parkes wide-field survey, and confirmed by a targeted ATCA followup
observation (spatial resolution $41''\times32''$, spectral resolution
3~\kms{}) (see Figure~\ref{fig:n1052_n5044}). This new group member,
designated J1320-1426, lies $\sim\degr{2.5}$ (1.4~Mpc) north-east of
the group centre and has velocity gradient north-west to south-east
centred at a velocity of 2750~\kms{}.  At the assumed distance we
derive an \hi{} mass of $\Mhi~=~1.3~\times~10^9$~\Msun{} and a
projected size of $35 \times 20$~kpc.  The DSS1 optical image of the
region reveals two peaks straddling the \hi{} peak, and a faint halo
of optical emission spatially coincident with the \hi{} source. This,
together with a surface density of $\sigma_{\mathrm{HI}} = 1.2 \Msun$
suggests that this may be a low surface brightness galaxy \cite{dmv}.
We have obtained optical images of J1320-1426, which will give us a
better understanding of its nature.

\section{IC~1459 Group}
The Parkes wide-field data reveal \hi{} emission within the X-ray
virial radius of the IC~1459 group, where extended, potentially
non-virialised, X-ray gas is known to exist. 
The centre of the group was imaged in ATCA followup observations
(spatial resolution $36'' \times 33''$, spectral
resolution~10~\kms{}) to determine the distribution and kinematics
of the \hi{} within the virial radius.
We detected \hi{} associated with the galaxies 
IC~5269B ($\Mhi = 1.1 \times 10^9~\Msun$), 
IC~5264 ($\Mhi = 0.2 \times 10^9~\Msun$) which exhibited two lobes of
emission, and NGC~7418A ($\Mhi = 1.6 \times 10^9~\Msun$) whose
\hi{} emission is tidally distorted.
In addition, three low \hi{}-mass clouds are tentatively detected near
IC~1459, IC~5264 and NGC~7418A. 
$\lambda$-21~cm radio continuum emission was also detected from IC~1459.

\section{NGC~3557 Group}
The Parkes survey observations of this group detected little \hi{} gas
outside the X-ray virial radius, but a substantial amount within this
region, where X-ray studies have detected 
complex emission.
The ATCA followup observation (spatial resolution $ 97''\times 60''$,
spectral resolution 7~\kms{}) detected \hi{} and radio continuum
emission from the galaxy NGC~3568 ($\Mhi = 3.8 \times 10^9~\Msun$).
\hi{} emission from NGC~3557\_17 \cite{zm00} was also detected
($\Mhi = 0.1 \times 10^9~\Msun$), and two tentative \hi{} detections
were made, near NGC~3557 and NGC~3557\_17. 

\begin{figure}
  \begin{minipage}[h]{0.45\linewidth}
    \begin{center}
      \resizebox{!}{7cm}{
        \includegraphics{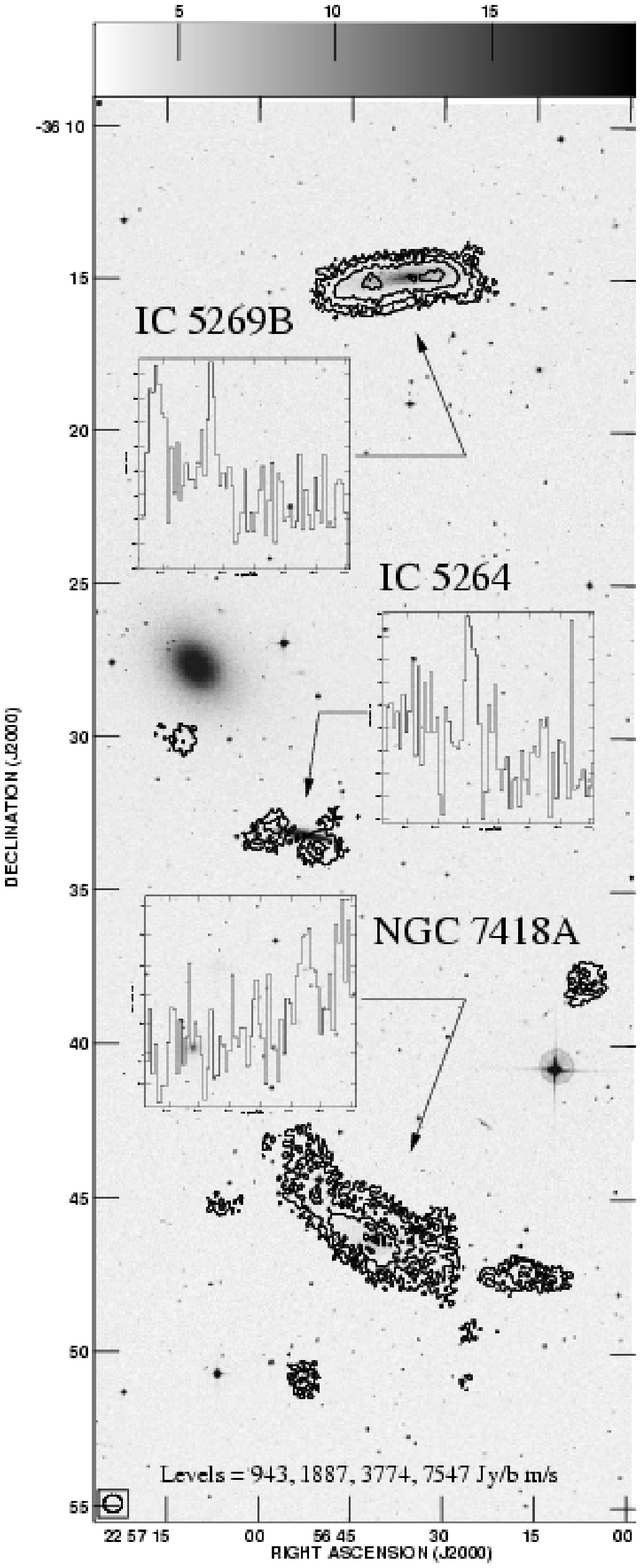}  
      }
    \end{center}
  \end{minipage} \hfill
  \begin{minipage}[h]{0.45\linewidth}
    \begin{center}
      \resizebox{\textwidth}{!}{
        \includegraphics{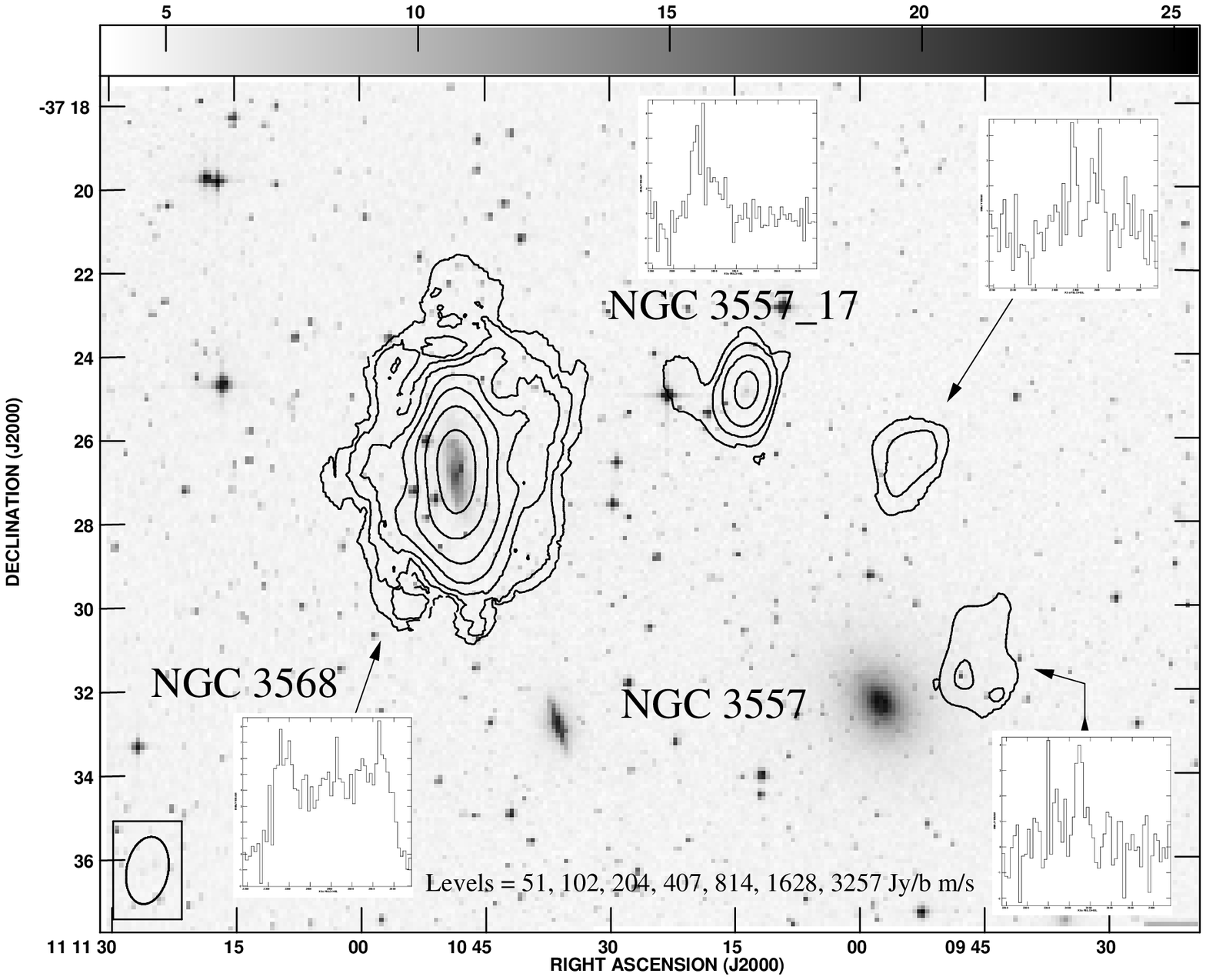}
      }
    \end{center}
  \end{minipage}

  \caption{ IC~1459 Group: \hi{} total intensity contours of IC~1459
  Group {\em (Left)} and NGC~3557 Group {\em (Right)}, superimposed on
  DSS2-red. The ATCA \hi{} spectra are shown next to the corresponding
  contours.}
  \label{fig:i1459_n3557}
\end{figure}

\section{Conclusion}
In combination with forthcoming results from the remaining 13 groups in
this study, and in parallel with the work being carried out on this
sample at X-ray and optical wavelengths, these results will help us
to understand how the neutral hydrogen gas in loose galaxy
groups, through its interactions with other group elements,
contributes to the evolution of the galaxies and the groups
they inhabit.

\bibliographystyle{klunamed}

\end{article}
\end{document}